\documentclass[a4paper,11pt]{article}
\usepackage{latexsym}
\usepackage{epsfig, subfigure, subfloat, graphicx, float }
\usepackage{amssymb}
\usepackage{amsmath}
\usepackage{epstopdf}
\usepackage{authblk}
\title{\bf Symmetron with a non-minimal kinetic term}  \author[1]{M. Honardoost
\thanks{m\_honardoost@sbu.ac.ir}}
\author[2]{D. F. Mota\thanks{D.F.Mota@astro.no}}
\author[1]{H. R. Sepangi \thanks{hr-sepangi@sbu.ac.ir}}
\affil[1]{\small Department of Physics, Shahid Beheshti University, G.C., Evin, Tehran
19839, Iran}
\affil[2]{\small  Institute of Theoretical Astrophysics, University of Oslo, 0315 Oslo, Norway }

\begin{document}

\maketitle
\begin{abstract}
We investigate the compatibility of the Symmetron with dark energy by introducing a non-minimal kinetic term associated with the Symmetron. In this new model, the effect of the friction term appearing in the  equation of motion of the Symmetron field becomes more pronounced due to the non-minimal kinetic term appearing in the action and, under specific conditions after symmetry breaking, the universe experiences an accelerating phase which,  in spite of the large effective mass of the scalar field, lasts as long as the Hubble time $H_{0}$.

\end{abstract}

\section{Introduction}

Many observations indicate that our universe today is in an accelerating phase
\cite{accel}. The standard $\Lambda \textit{CDM}$ describes late time positive acceleration via a mysterious component known as the Cosmological Constant (CC) $\Lambda$. Although standard $\Lambda CDM$ has passed all gravity tests in linear regimes and agrees well with planck data \cite{planck}, clarifying the nature of $\Lambda$ (old CC problem) is still an open problem. The vacuum energy is the first candidate for $\Lambda$ but when we properly consider the renormalized vacuum energy of particles in the Standard Model we find that the theoretical value is about 55 order of magnitude greater than the observational value \cite{Martin}. Apart from the old problem, the new one which refers to why we are in an accelerating phase today, is another challenge for the standard $\Lambda CDM$ \cite{yashar}.

As a solution to the cosmological problems, modified gravity was proposed and attracted considerable
attention \cite{modgrav}. In this new framework, one can either modify the geometry part of General Relativity (GR) or alternatively add an additional, usually a scalar field, which is known as  ``dark energy.'' In this regard, hypothetical dynamical scalar fields
such as Quintessence have been postulated  to explain the observed
acceleration of the universe \cite{caldwell}. However, the important question is: if the universe contains an extra scalar field why the effects of the fifth force associated with this extra degree of freedom has not yet been detected? In other words, although  GR has not been tested directly at large scales, we are almost sure that it works accurate enough in linear regimes, e.g. in the Solar System \cite{yashar}. So if we propose any modified theory of gravity, the new model must reduce to GR for local observations. This problem exists for any extra scalar field and one of the innovative proposals to tackle it is to consider screening mechanisms \cite{brax}. By means of screening, the scalar field participates in a mechanism which makes it locally hidden from
gravitational observations while its effects  may be observed  at cosmological scales. Chameleon \cite{chamsym}, Vainstein \cite{vainstein}, Symmetron \cite{symcos} and Dilaton \cite{braxdil} are the most significant screening models which have been introduced so far. For these models the screening mechanism occur  through the coupling between scalar and baryonic/dark matter which let the scalar field to be concealed when the environment is dense but revealed when the environment is rare in comparison to a critical value. Several local and cosmological tests for such models have been investigated \cite{mota1,mota2,mota3,mota4}. Among the many possible screening mechanisms and models, in this work we are interested in the Symmetron since it uses symmetry breaking of the effective potential to participate in screening. The symmetry breaking mechanism was used in \cite{linde} to present an inflationary model and even inspired people to explain the late time behavior of the universe \cite{symcos,mypap1,thawing}.

However, in the original Symmetron model \cite{symcos}, local gravity
experiments impose constraints on parameters of the model which make the effective mass of the scalar field too large. Hence, immediately after symmetry breaking the scalar field falls to a new minimum. Thus if the accelerating phase of the universe begins at the onset of the symmetry breaking,  it will not last long enough, unless one fine tunes the height of the Symmetron potential minima to a non-zero value corresponding to a cosmological constant.
Although it is shown in \cite{mypap1} that the problem of the original Symmetron is not limited to what was said above, in this paper we shall modify the Symmetron so as to get an acceleration for a time period of the order of the Hubble time after symmetry breaking. To do so, inspired by \cite{infl}, we propose Symmetron with a non-minimal derivative coupling (NMDC) term. As it was shown in \cite{germani} in the contex of ``gravitationally enhanced friction" \cite{gef}, this new term makes the Symmetron field to roll down the potential as slow as is necessary.

The non-minimal coupling between the scalar field and Ricci scalar is well-known as in Brans-Dicke models in the Jordan frame \cite{burin}, or the non-minimal scalar derivative coupling with the Ricci scalar proposed in \cite{amen} by Amendola. Later on, Capozziello, Lambiase and Schmidt studied the cosmology of the Einstein-Hilbert action containing $R\phi _{,\mu}\phi^{,\mu}$ and $R_{\mu \nu}\phi_{,\mu}\phi^{,\nu}$ together with a free kinetic term and a scalar potential \cite{calasch, capolam}. As a novel effort to accommodate quantum gravity perturbatively, Granda proposed the two coupling constants NMDC model where in this new proposal, the usual Einstein-Hilbert action including matter contains the minimal scalar kinetic term, scalar potential and augmented by two NMDC terms, $-\frac{1}{2} \kappa R \phi^{-2} g_{\mu \nu} \phi^{,\mu}\phi^{\nu}$ and $-\frac{1}{2}\eta \phi^{-2} R_{\mu \nu} \phi^{,\mu}\phi^{\nu}$,  where the dimensionless couplings $\kappa $ and $\eta $ are re-scaled by $\frac{1}{\phi^{2}}$ \cite{granda}.
Among these models we are interested in Sushkov's in which a special case of the Granda model is considered where the coupling constants are proportional to each other, thus the combination includes the Einstein tensor $\kappa G_{\mu \nu}\phi^{,\mu}\phi^{\nu}$  \cite{sushkov}. Fortunately, adding such a non-minimal kinetic term keeps our model in the Horndeski
fields category if we take $G_{3}=0$ in $L_{3}$, $G_{4}=1$ in $L_{4}$ and $G_{5}=1$ in
$L_{5}$ \cite{ppn,hornd}, so we can be sure to have a ghost free model and that the field equations will remain second order differential equations.

In this paper we start by briefly reviewing the Symmetron model. In section 3 we introduce the action  with non-minimal kinetic term and consider the field equations in both Newtonian and Post-Newtonian limits to see if any  constraints are imposed on the parameters of the model. Finally in section 4 we study the cosmological behaviour of the model and show that it can lead to late time acceleration under specific conditions. Finally, using a numerical example we show that the duration of the accelerating time can be as long as the Hubble parameter $H_{0}$.

\section{The Symmetron}
Although we can look at the Symmetron as a class of Chameleon field, in the Symmetron model, it is the vacuum expectation value of the scalar field, $\phi_{VEV}$, which depends on the local environment. In other words, $\phi_{VEV}$  takes two different values where density of the environment changes in such a way as to make the expectation value zero where the local density is high,  while $\phi_{VEV}$ becomes non-zero where the local density is low.
Screening in this model occurs because detectable fluctuations $\delta \phi$ couple to density which is proportional to $\phi_{VEV}$. Thus, whenever $\phi_{VEV}=0$, e.g. in local experiments, the matter decouples from the scalar field while for $\phi_{VEV}\neq 0$ the scalar effects, as a result of the scalar-matter coupling, could be detectable.
In this section we briefly review  the original Symmetron and discus  the failures of the model as
a candidate for dark energy.

The original Symmetron model (or Symmetron with non-minimal kinetic term) is based on the action
\begin{equation}\label{eq1}
  S=\int d^{4}x \sqrt{-g} \left[\frac{M^{2}_{pl}}{2}R -\frac{1}{2}(\nabla \phi)^{2}-V(\phi)\right]+
  S_{m}[A^{2}(\phi)g_{\mu \nu},\Psi_{m}],
\end{equation}
where $M_{pl}$ is the reduced Planck mass and $S_{m}$ refers to the action of any matter field $\Psi_{m}$ and both the scalar potentials $V(\phi)$ and $A(\phi)$ must have $\mathbb{Z}_{2}$ symmetry. As an example one may choose
\begin{eqnarray}
   && V(\phi)=-\frac{1}{2}\mu^{2}\phi^{2}+\lambda\frac{1}{4}\phi^{4}\label{eq2}, \\
   && A(\phi)=1+\frac{\phi^{2}}{2M^{2}}\label{eq3},
\end{eqnarray}
where $\mu^{2}$ and $M$ are mass scales of the model and $\lambda$ is a positive dimensionless coupling. Variation of (\ref{eq1}) with respect to $g_{\mu \nu}$ and $\phi$ leads to equations of motion
\begin{eqnarray}
   && \Box \phi = \frac{dV(\phi)}{d\phi}-A^{3}(\phi)\frac{dA(\phi)}{d\phi}\tilde{T}\label{eq4}, \\
   && M^{2}_{pl}G_{\mu \nu}= T^{\phi}_{\mu \nu}+ A^{2}(\phi)\tilde{T}_{\mu \nu}\label{5},
\end{eqnarray}
where $\tilde{T}=\tilde{g}^{\mu \nu}\tilde{T}_{\mu \nu}$ is the trace of the matter energy-momentum
tensor in the Jordan frame. By substituting (\ref{eq2}) and (\ref{eq3}) in (\ref{eq4}) one can see that the effective potential for dust takes the form
\begin{equation}\label{eq6}
  V_{eff}=\frac{1}{2}\left(\frac{\rho}{M^{2}}-\mu^{2}\right)\phi^{2}+\frac{\lambda}{4}\phi^{4},
\end{equation}
for which $\rho\equiv A^{3}(\phi)\tilde{\rho}$ \cite{symsym}.
According to (\ref{eq6}), it is clear that the effective mass $m^{2}_{eff}=\frac{\rho}{M^{2}}-\mu^{2}$ depends on local density of the environment, so that for $\rho > \rho_{crit}=M^{2}\mu^{2}$ the effective potential is parabolic around $\phi_{EVE}=0$ while for $\rho< \rho_{crit}=M^{2}\mu^{2}$, it becomes tachyonic so that the symmetry of the effective
potential is broken and $\phi_{EVE}$ tends to a new minimum, $\phi_{EVE}= \pm
\sqrt{(\mu^{2}-\frac{\rho}{M^{2}})/\lambda}$, from its initial zero value. By local tests of gravity, constraints on the parameters of the model can be found. As was shown in \cite{symcos}, local gravity constraints lead to an upper bound on $M$ such that $M\leq10^{-4}M_{pl}$. From Symmetron model we demand that symmetry breaking occurs at red shifts $z\sim 1$. This requirement sets the value of the critical potential to $\rho_{crit}\simeq\rho_{0}=M^{2}_{pl}H^{2}_{0}$ which consequently leads to $\mu^{2}\simeq\frac{H^{2}_{0}M^{2}_{}pl}{M^{2}}$. Hence after symmetry breaking, mass of the scalar field is $m_{s}=\sqrt{2}\mu\simeq\frac{H_{0}M^{2}_{}pl}{M^{2}}$. By considering an upper bound on $M$ one can easily see that mass of the scalar field becomes $m_{s}\geq10^{4}H_{0}$ which is too large to play the role of dark energy. Thus, as a result of a large mass in the minimal case, the scalar field rolls down too fast towards the new non-zero minimum.
It therefore seems necessary to find a way to slow it down, for example by
creating or increasing a friction term.
Looking at (\ref{eq4}), we find that the box term on the left-hand side includes a friction-like term and since it is a consequence of the kinetic term in action (\ref{eq1}), it seems reasonable to assume that the appearance of Symmetron with non-minimal kinetic term would make the scalar field to roll down slowly enough after the symmetry breaking.

In the following section we first check if local constraints on parameters in the new model would be altered, followed by focusing on the cosmological behaviour and the effects of symmetry breaking on the scalar field.

\section{Newtonian and Post-Newtonian approximations}

In this section we follow \cite{ppn} to check if the Newtonian or Post-Newtonian approximation of the proposed model leads to different constrains on the parameters as compared to that of the Symmetron.

As was mentioned before, to avoid the problems of the Symmetron model we use the action of the Symmetron with a non-minimal kinetic term
\begin{equation} \label{7}
S=\int
d^{4}x\sqrt{-g}\Big\{\frac{R}{16\pi
G}-\frac{1}{2}\left[g_{\mu\nu}-\omega^{2}G_{\mu\nu}\right]\nabla^{\mu}\phi\nabla^{\nu}\phi-V(\phi)\Big\}
+S_{m}[\widetilde{g}_{\mu\nu},\Psi_{m}],
\end{equation}
where $\Psi_{m}$ indicates all possible baryonic/dark matter fields, with the coupling constant $\omega^{2}$ having dimension $(\mbox{length})^{2}$. By varying the above action with respect to $g_{\mu\nu}$, the equation of motion takes the form
\begin{equation}\label{8}
\frac{R_{\mu\nu}}{8\pi G}= S^{(m)}_{\mu\nu}+S^{(\phi)}_{\mu\nu}-\omega^{2}S^{(\theta)}_{\mu\nu},
\end{equation}
where $S_{\mu\nu}=T_{\mu\nu}-\frac{1}{2}g_{\mu\nu}T$, $S^{(m)}_{\mu\nu}$ corresponds to the matter part of the action which we have considered as a perfect fluid, $S^{(\phi)}_{\mu\nu}$ represents the minimal kinetic derivative part of the scalar field and
$S^{(\theta)}_{\mu\nu}$ results from the coupled scalar field kinetic term \cite{infl}
\begin{equation}\label{9}
  S^{(\phi )}_{\mu\nu}=\nabla_{\mu}\phi \nabla_{\nu}(\phi)+g_{\mu\nu}V(\phi),
\end{equation}
\begin{eqnarray}\label{10}
  && S^{(\theta)}_{\mu\nu}= - \frac{1}{2}\nabla_{\mu}\phi\nabla_{\nu}\phi R
  +2\nabla_{\alpha}\phi\nabla_{(\mu}\phi R^{\alpha}_{\nu )}
    -\frac{1}{2}(\nabla\phi)^{2}G_{\mu\nu}+\nabla^{\alpha}\phi\nabla^{\beta}\phi
R_{\mu\alpha\nu\beta}
   +\nabla_{\mu}\nabla^{\alpha}\phi\nabla_{\nu}\nabla_{\alpha}\phi
   \nonumber \\
   && -\nabla_{\mu}\nabla_{\nu}\phi\Box\phi-\frac{1}{2}g_{\mu\nu}\nabla_{\alpha}\phi\nabla_{\beta}\phi R^{\alpha\beta}.
\end{eqnarray}
Similarly, equation of motion of the scalar field becomes

\begin{equation}\label{11}
[g^{\mu\nu}-\omega^{2}G^{\mu\nu}]\nabla_{\mu}\nabla_{\nu}\phi-V'(\phi)+A'(\phi)\rho=0,
\end{equation}
where a prime indicates derivative with respect to $\phi$ and $\rho\equiv A^{-1}(\phi)\rho_{E}$
is an un-physical $\phi$-independent quantity which we also deal with in the Symmetron model  \cite{symsym} and $\rho_{E}$ is total energy density in the Einstein frame.
\subsection{Newtonian approximation}
Before dealing with perturbations to obtain the Newtonian or Post-Newtonian limits, we need to clearly justify our assumptions. Following \cite{ppn} we assume that the background scalar field $\phi_{0}$ is homogeneous and only varies with cosmological time scale, so that $\phi_{0}$ is constant in the solar system. Also, the
background is taken as an empty universe which is characterized by the Minkowski metric $\eta_{\mu\nu}$ and constant  scalar field $\phi_{0}$. Taking the matter field  as the static gravitational source in this background leads to linear  perturbations in both metric and scalar field according to
\begin{eqnarray}\label{12}
&&g_{\mu\nu}= \eta_{\mu\nu}+h_{\mu\nu}, \nonumber \\
&&\phi = \phi_{0}+\delta\phi,
\end{eqnarray}
Thus equations (\ref{eq4}) and (\ref{5}) become
\begin{equation}\label{13}
  \frac{\delta R_{\mu\nu}}{8\pi G}=\delta S^{(m)}_{\mu\nu}+\delta S^{(\phi)}_{\mu\nu}-\omega^{2}\delta
  S^{(\theta)}_{\mu\nu},
\end{equation}
and
\begin{equation}\label{14}
  \eta_{\mu \nu}\delta(\nabla_{\mu}\nabla_{\nu}\phi)-\delta(V'(\phi)-A'(\phi)\rho)=0.
\end{equation}
Note that we are perturbing our dynamical field in terms of planetary velocity $v^{2}$ \cite{will} and by $n$-th order perturbation we mean that we are interested in having time-like geodesic equations up to terms $(v^{2})^{n}$.

The value of $\phi_{0}$ would be determined by considering the zeroth order perturbation.
Thus Equation (\ref{13}) takes the form
\begin{equation}\label{15}
  \eta_{\mu\nu}V(\phi_{0})=0,
\end{equation}
while to zeroth order, from (\ref{14}) we have
\begin{equation}\label{16}
  -V'(\phi_{0})+ A'(\phi_{0})\rho=0,
\end{equation}
with a solution given by $V'(\phi_{0})=0$ and $A'(\phi_{0})=0$. Equation (\ref{15}) together with (\ref{16}) indicate that in the
background, $\phi_{0}$  makes the potential and its derivative to vanish \cite{ppn}. Thus we are inclined to expand the scalar field around this special point, otherwise expanding around a different point will break the consistency of zeroth order perturbation. Note that in any Symmetron-like screening model,  one  can choose the potentials $V(\phi)$  and  $A(\phi) $ so that the effective potential has a minimum. Hence we are practically expanding the scalar field around the minimum (stable point) of the effective potential and therefore a perturbative expansion around any other point is not valid.  Consequently, the first order perturbation or the Newtonian limit would be as useful as the zeroth order.

Since we have assumed non-relativistic and quasi-static approximation for gravitational fields \cite{hohmann,barreira}, then to first order in perturbation  from equation (\ref{14}) we have
\begin{equation}\label{17}
  \nabla^{2}\delta\phi=(V''_{0}-A''_{0}\rho^{(0)})\delta\phi.
\end{equation}
We should emphasise at this point that in order to act as dark energy,  the scalar field is assumed to evolve at scales comparable to cosmic time so that at scales much smaller, i.e. solar scales, we can ignore the time derivative of fluctuation $\delta \phi$ and just focus on $ \delta \phi (\vec{x})$. Therefore, in this {\it quasi-static} approximation \cite{barreira}, one is allowed to ignore the time derivative of $\delta\phi$ in favour of the spatial derivative of the scalar field.    We also note that in our model the scalar field evolves even slower due to the non-minimal kinetic coupling. Thus in this and next section we ignore the time derivative of $\delta \phi$ compared to its spatial derivative \cite{hohmann}. Although the deviation from quasi-static approximation is not the main concern in solar scales, there are some arguments in cosmological perturbation theory literature  which focus on the effects of spatial scales larger then $\sim10 h^{-1}Mpc$ \cite{mota,pedro}  in $N$-body simulations and structure formation.

As is shown in \cite{ppnscreen}, one can solve equation (\ref{17}) for suitable boundary conditions \cite{chamsym} and find $\delta\phi$ to order ${\cal O}(\delta\phi)\sim 1$.
Remembering that as we have taken $A'(\phi_{0}=0)$, equation (\ref{17}) is consistent with the scalar field equation of motion in the Symmetron model for the outer solution of a static source in the Newtonian limit \cite{symcos}. Also, since we have considered $\phi_{0}$ to be homogeneous and $G^{(0)}_{\mu \nu}=0$ , this agreement is reasonable and we do not expect to see any trace of the non-minimal term in (\ref{17}) for the Newtonian limit.

To keep terms of order one in (\ref{13}) we must first compute the zeroth order components of $ \delta
S^{(m,0)}_{\mu\nu}$ but first order components of $ \delta S^{(\phi,1)}_{\mu\nu}$ and $ \delta S^{(\theta,1)}_{\mu\nu}$

\begin{eqnarray}\label{18}
&&\delta S^{(m,0)}_{00}=\frac{1}{2}\rho^{(0)}_{E}, \nonumber \\
&&\delta S^{(m,0)}_{ij}=\frac{1}{2}\delta_{ij} \rho^{(0)}_{E}, \nonumber \\
&&\delta S^{(m,0)}_{0i}=\delta S^{(\phi,1)}_{\mu\nu}=\delta S^{(\theta,1)}_{\mu\nu}=0.
\end{eqnarray}
Now, using
\begin{eqnarray}\label{19}
&& \delta R^{(1)}_{00}=\frac{1}{2}\nabla^{2}h^{(1)}_{00}, \nonumber \\
&& \delta R^{(1)}_{ij}=\frac{1}{2}\nabla^{2}h^{(1)}_{ij},
\end{eqnarray}
which are  written in harmonic gauge \cite{will}, one can keep terms up to order one in both sides of (\ref{13}) to obtain
\begin{eqnarray}\label{20}
  && \nabla^{2}h^{(1)}_{00}= -8\pi G A_{0}\rho^{(0)}, \nonumber \\
  && \nabla^{2}h^{(1)}_{ij}= -8\pi G \delta_{ij}A_{0}\rho^{(0)},
\end{eqnarray}
which indicate that $ g_{\mu\nu}$ is not affected by the non-minimal kinetic term in the Newtonian limit.
We therefore conclude that in this limit, the equations of motion in the original Symmetron model are similar to field equations in the Symmetron model with non-minimal kinetic term and no gravity test is able to distinguish between them in {\it{Newtonian approximation}}.

\subsection{Post-Newtonian approximation}
Since in Post-Newtonian approximation we take time-like geodesic equations up to order two,  we
need to compute some (instead of all) components of the metric up to this order. It is therefore sufficient to only calculate
$h^{(1)}_{00}$ ,$h^{(2)}_{00}$ (in term of $ v^{2} $ and $ v^{4}$), $h^{(1.5)}_{0i}$  (in terms of $v^{3})$ and finally $h^{(1)}_{ij}$ (in terms of $v^{2}$) in the harmonic or any other suitable gauge.

In what follows  we seek to see if our proposed model behaves differently in Post-Newtonian limit compared to that of the Symmetron. Provided that our results show no difference, we can conclude that no additional constraints from local tests of gravity would be imposed on our model.
Now, keeping terms to second order  in (\ref{13}), we have
\begin{eqnarray}\label{21}
   && \frac{\delta R^{(2)}_{00}}{8\pi G}=\delta S^{(m,1)}_{00}+\delta S^{(\phi,2)}_{00}-\omega^{2}\delta
   S^{(\theta,2)}_{00}, \nonumber \\
   && \frac{\delta R^{(1.5)}_{0i}}{8\pi G}=\delta S^{(m,0.5)}_{0i}+\delta
   S^{(\phi,1.5)}_{0i}-\omega^{2}\delta S^{(\theta,1.5)}_{0i}, \nonumber \\
   && \frac{\delta R^{(1)}_{ij}}{8\pi G}=\delta S^{(m,0)}_{ij}+\delta S^{(\phi,1)}_{ij}-\omega^{2}\delta
   S^{(\theta,1)}_{ij}.
\end{eqnarray}
Solutions of the above equations provide us with the desired components of $h_{\mu \nu}$ to the intended order. However,  for our purpose here, rather than solving the above equations explicitly we can directly check to find if they are the same as that of the original Symmetron model. Going back to (\ref{13}), we immediately realize that $\delta S^{(\theta)}_{\mu\nu}$ is the only term which would make any difference, so we have to compute $\delta S^{(\theta ,1.5)}_{i0}$ and $\delta S^{(\theta ,2)}_{00}$
to compare $\delta R^{(1.5)}_{0i}$ and $\delta R^{(2)}_{00}$ in both models (we note that $\delta S^{(\theta ,1)}_{ij}=0$, as shown above).
Now, perturbing (\ref{10}) and keeping terms up to order 1.5 and 2 we obtain

\begin{eqnarray}\label{22}
&&\delta S^{(\theta,1.5)}_{\mu\nu}=0, \nonumber\\
&&\delta
S^{(\theta,2)}_{\mu\nu}=\partial_{\mu}\partial^{i}\delta\phi\partial_{\nu}\partial_{i}\delta\phi-
(\partial_{\mu}\partial_{\nu}\delta \phi)\nabla^{2} \delta \phi.
\end{eqnarray}
It is interesting to note that in quasi-static approximation, the $\partial_{t}\delta \phi$ term is ignored while the relation $\delta S^{(\theta ,2)}_{00}=0$ is drived which shows that no trace of the non-minimal kinetic term can be found neither in $\delta R^{(1.5)}_{i0}$ and $\delta R^{(2)}_{00}$  nor in $\delta R^{(1)}_{00}$ and $\delta R^{(1)}_{i0}$ as we have seen in the previous subsection. So we conclude that \textit{the Symmetron and Symmetron with non-minimal kinetic term have the same behavior in both
the quasi-static Newtonian and  Post-Newtonian limits and that there is no difference between constraints from local tests of gravity in both models. Therefore the mass of the scalar field in the non-minimal model is} $ m_{\phi}\geq 10^{4}H_{0}$, \textit{just like in the Symmetron model.}
However,  the difference may be detected in the time evolution of the scalar field in cosmological scales where the non-minimal term makes the scalar field to roll down slowly. In the next section we will discus cosmological aspect of the model.

\section{Cosmology of Symmetron with non-minimal kinetic term}

In this section our purpose is to investigate the possibility of our universe, filled with matter and scalar fields with a non-minimal kinetic term, to have accelerated expansion via symmetry breaking, specially if  such acceleration lasts as long as the Hubble time.
Although the Symmetron (in the case of minimal kinetic term) is used to explain late time acceleration via symmetry breaking,  it has not been successful in this scenario because as soon as the matter content density of the universe becomes less than $\rho_{crit}$, the scalar field becomes massive and rolls down too fast to the new minimum and oscillates around it, causing the accelerating phase to end. Acceleration with non-minimal kinetic term through slow roll condition with positive potential has been studied before in \cite{warminf} but our attempt here is to check for the positive acceleration in the case of negative potential according to action (\ref{7}) and potential (\ref{eq2}).
Returning to action (\ref{7}) and considering a homogeneous and isotropic universe which is described by the FRW metric
\begin{equation}\label{23}
  ds^{2}=-dt^{2}+a^{2}(t)(dr^{2}+r^{2}d\Omega^{2}),
\end{equation}
where $a(t)$ is the scale factor and $d\Omega^{2}=d\theta^{2}+{\sin^{2}(\theta)}d\phi^{2}$, one can derive
Friedmann equations, continuity equation for the conserved quantity $\rho=A^{-1}(\phi)\rho_{E}$ and the scalar field equation of motion as follows

\begin{eqnarray}
   && H^{2}=\frac{8\pi G}{3}(\frac{1}{2}\dot{\phi}^{2}+
   V(\phi)+\frac{9}{2}\omega^{2}H^{2}\dot{\phi}^{2}+\rho_{E})\label{24}, \\
   && \dot{H}=-4\pi G \dot{\phi}^{2}+4\pi G\omega^{2}(\dot{H}-3H^{2}+H\ddot{\phi}\dot{\phi}^{-1})-4\pi
   G\rho_{E}\label{25}, \\
   && \dot{\rho}+3H\rho=0\label{26}, \\
   && (\ddot{\phi}+3H\dot{\phi})+3\omega^{2}(H\ddot{\phi}+2H\dot{H}\dot{\phi}+3H^{3}\dot{\phi})=
   -V_{eff,\phi}\label{27}.
\end{eqnarray}
In a similar fashion to Symmetron,  we require that:\newline

I- Before symmetry breaking and $t<t_{crit}$, the universe is in a decelerating phase.
\newline

II- After symmetry breaking and $t>t_{crit}$, the universe returns to an accelerating phase.\newline

III- To detect the acceleration today, the accelerating phase must last for a time of order of the Hubble time.\newline
\newline
The sign of $\ddot{a}(t)$ can be read from
\begin{eqnarray}\label{28}
    \frac{\ddot{a}(t)}{a(t)}&=& H^{2}+\dot{H}= \nonumber\\
   &&  \frac{8\pi G}{3}\frac{\frac{1}{2}\dot{\phi}^{2}+V(\phi)+\rho_{E}}{1-12\pi G\omega^{2}\dot{\phi}^{2}} +
   \frac{-4\pi G[\dot{\phi}^{2}+3\omega^{2}H^{2}\dot{\phi}^{2}-\omega^{2}H\ddot{\phi}\dot{\phi}]-4\pi
   G\rho_{E}}{1-4\pi G\dot{\phi}^{2}\omega^{2}}.
\end{eqnarray}
In the case of a minimal kinetic term where $\omega^{2}=0$, relation (\ref{28}) reduces to
\begin{equation}\label{29}
  \frac{\ddot{a}}{a}=\frac{8\pi G}{3}\left(-\frac{1}{2}\dot{\phi}^{2}+V(\phi)-\frac{1}{2}\rho_{E}\right).
\end{equation}
 During $t<t_{crit}$, the effective potential
 $V_{eff}\cong\frac{1}{2}\frac{\rho}{M^{2}}\phi^{2}+\frac{\lambda}{4}\phi^{4}$ is parabolic around
 $\phi_{min}=0$ for both minimal or non-minimal kinetic models and hence by setting $\phi_{0}=0$ as the initial  condition,  equation (\ref{28}) takes the following form
  \begin{equation}\label{30}
  \frac{\ddot{a}}{a}=-\frac{4\pi G}{3}\rho_{E}<0,
\end{equation}
which is consistent with the sign of $\frac{\ddot{a}}{a}$ before symmetry breaking. However, (\ref{29}) is clearly negative for $V(\phi)<0$ and we therefore conclude that a negative potential cannot lead to positive acceleration in the case of minimal kinetic term even after symmetry breaking \cite{mypap1}. From (\ref{28}), it is clear that the conclusion is not valid for negative potentials  in the case of non-minimal kinetic term and under specific conditions the
universe goes over to an accelerating phase as we will see in what follows.

For $\omega^{2}\neq0$, as the time passes and matter density of the universe decreases,  the universe
arrives at $t=t_{crit}$ where $\rho=\rho_{crit}=\mu^{2}M^{2}$. The squared effective mass $ m_{eff}^{2}$ then becomes negative and the effective potential takes the form $V_{eff}\cong
-\frac{1}{2}\mu^{2}\phi^{2}+\frac{\lambda}{4}\phi^{4}$ approximately. Hence the initial point $\phi_{0}=0$ is not stable any more and the scalar field becomes a dynamical field, now having a non-zero minimum. If we consider the slow-roll conditions
\begin{eqnarray}
   && \ddot{\phi}\ll\dot{\phi}H\label{31}, \\
   && \dot{\phi}^{2}\ll |V(\phi)|<\frac{9}{2}\omega^{2}H^{2}\dot{\phi}^{2}\label{32},
\end{eqnarray}
then we can approximate (\ref{28}) as
\begin{equation}\label{33}
\frac{\ddot{a}}{a}\simeq\frac{8\pi G}{3}\frac{V(\phi)+\rho_{E}}{1-12\pi G\dot{\phi}^{2}\omega^{2}} +
\frac{-4\pi G[\dot{\phi}^{2}+3\omega^{2}H^{2}\dot{\phi}^{2}+\rho_{E}]}{1-4\pi G\dot{\phi}^{2}\omega^{2}},
\end{equation}
where we have used $\dot{\phi}\ddot{\phi}H\ll\dot{\phi}^{2}H^{2}$ for $\dot{\phi}H>0$. It is clear
that the first fraction in (\ref{33}) indicates $H^{2}$ while the second is $\dot{H}$ and thus to have an accelerating phase after symmetry breaking  we need
\begin{eqnarray}\label{34}
  &&    \dot{H}>0, \nonumber \\
  \mbox{or} &&   \dot{H}<0\hspace{3mm}   \mbox{and} \hspace{3mm}   H^{2}>|\dot{H}|.
\end{eqnarray}
For the case $|V(\phi)|>\rho_{E}$, the positivity of $H^{2}$ imposes
\begin{equation}\label{35}
  1<12\pi G\omega^{2}\dot{\phi}^{2}.
\end{equation}
From the second fraction in (\ref{33}) it is seen that $\dot{H}>0$ leads to
\begin{equation}\label{36}
  1<4\pi G \dot{\phi}^{2}\omega^{2}.
\end{equation}
However the positivity of $H^{2}$ together with $\dot{H}<0$ and $H^{2}>|\dot{H}|$ lead to
\begin{equation}\label{37}
  \frac{1}{3}<4\pi G \dot{\phi}^{2}\omega^{2}<1.
\end{equation}
From (\ref{35}), (\ref{36}) and (\ref{37}) we conclude that for $|V(\phi)|>\rho_{E}$ after symmetry breaking, the universe experiences an accelerating phase with $\dot{H}>0$ and condition (\ref{36}) but if $\dot{H}<0$, for interval (\ref{37}), positive acceleration will occur.
In the case of $|V(\phi)|<\rho_{E}$, the positivity of $H^{2}$ gives
\begin{equation}\label{38}
  1>12\pi G\omega^{2}\dot{\phi}^{2}.
\end{equation}
In this case when $\dot{H}>0$ the denominator of the second fraction in (\ref{33}) must be negative which yields
\begin{equation}\label{39}
  3<12\pi G \dot{\phi}^{2}\omega^{2},
\end{equation}
while $\dot{H}<0$ with $H^{2}>|\dot{H}|$ leads to
\begin{equation}\label{40}
  1 < 12\pi G\dot{\phi}^{2}\omega^{2} < \frac{3}{2}.
\end{equation}
As equations (\ref{39}) and (\ref{40}) are inconsistent with (\ref{38}) we conclude that in the case of $|V(\phi)|<\rho_{E}$ the positive acceleration does not occur for symetron with non-minimal kinetic term. What we have discussed above shows that after symmetry breaking in our model if the scalar field rolls down slowly according to (\ref{31}) and (\ref{32}), the accelerating phase of the universe can just begin  for a certain range of $4\pi G\dot{\phi}^{2}\omega^{2}$; when $4\pi G\dot{\phi}^{2}\omega^{2}>1$ the universe experiences an accelerating phase with $\dot{H}>0$ while $\frac{1}{3} < 4\pi G\dot{\phi}^{2}\omega^{2} < \frac{1}{2}$ leads to a positive acceleration with $\dot{H}<0$.

Let us now compare the evolution of the rescaled-scalar field via a specific example for both minimal and non-minimal kinetic terms. In the case of the Symmetron model and $\omega =0$, numerical solution of (\ref{24}-\ref{27}) for initial values $\tilde{\phi_{0}}=10^{-10}$, $\tilde{H_{0}}=2\sqrt{3}/3$, $\tilde{\rho_{0}}=4$ and dimensionless parameters $\tilde{M}= 10^{-4}$, $\tilde{\mu}=2\times 10^{4}$ and $\lambda \frac{H^{2}_{0}}{M^{2}_{pl}}=10^{20}$ leads to figure 1 which shows that just after symmetry breaking, the scalar field has settled down in a new non-zero minimum.
\begin{figure}[H]
\centering\epsfig{file=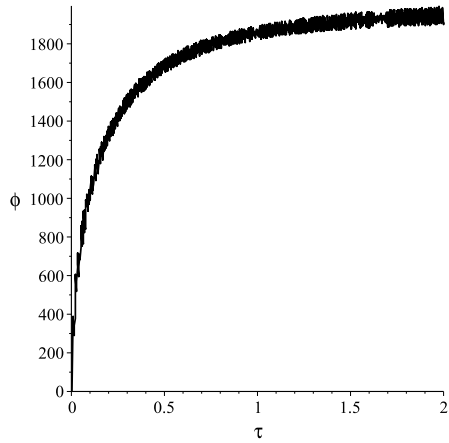,width=6cm,angle=0}
\caption{\footnotesize A schematic illustration of the rescaled scalar field for the Symmetron model ($\omega =0$) in terms of dimensionless time $\tau$ for $\tilde{\phi_{0}}=10^{-10}$, $\tilde{H_{0}}=2\sqrt{3}/3$, $\tilde{\rho_{0}}=4$, $\tilde{M}= 10^{-4}$, $\tilde{\mu}=2\times 10^{4}$ and $\lambda\frac{H^{2}_{0}}{M^{2}_{pl}}=10^{20}$ } \label{fig1}
\end{figure}
However, the evolution will change in the case of non-minimal kinetic term with $\omega = 10^{10}$ and the same initial conditions which is shown in figure 2.
\begin{figure}[H]
\centering\epsfig{file=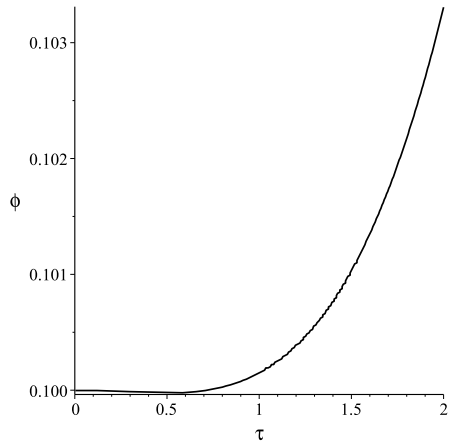,width=6cm,angle=0}
\caption{\footnotesize The schematic illustration of the Symmetron with non-minimal kinetic term and $\omega= 10^{10}$ model in terms of dimensionless time $\tau$ for $\tilde{\phi_{0}}=10^{-10}$, $\tilde{H_{0}}=2\sqrt{3}/3$, $\tilde{\rho_{0}}=4$, $\tilde{M}= 10^{-4}$, $\tilde{\mu}=2\times 10^{4}$ and $\lambda\frac{H^{2}_{0}}{M^{2}_{pl}}=10^{20}$} \label{fig2}
\end{figure}
From figures 1 and 2 it is clear that in the case of non-minimal kinetic term, the scalar field will arrive at the new minimum, taking longer  than that in the Symmetron model.
Finally, as a further support to our results, we have presented time derivative of the scalar field for both $\omega =0$ and $\omega=10^{10}$ cases in figures 3 and 4.
\begin{figure}[H]
\centering\epsfig{file=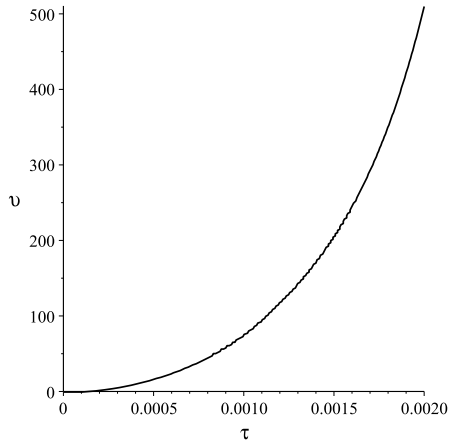,width=6cm,angle=0}
\caption{\footnotesize A schematic illustration of the time derivative of  rescaled scalar field $\nu\equiv\frac{d\tilde{\phi}}{d\tau}$ for the Symmetron model ($\omega =0$) in terms of dimensionless time. The initial conditions are the same as that in Figure 1. } \label{fig3}
\end{figure}
\begin{figure}[H]
\centering\epsfig{file=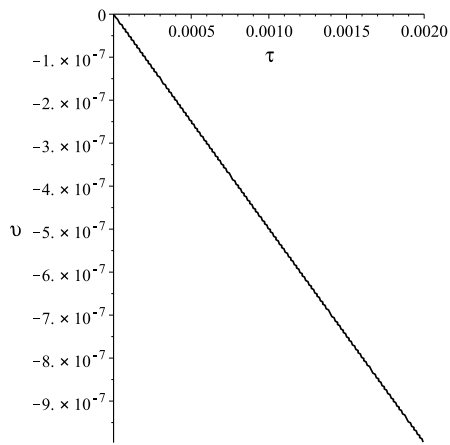,width=6cm,angle=0}
\caption{\footnotesize The schematic illustration of the time derivative of rescaled scalar field $\nu\equiv\frac{d\tilde{\phi}}{d\tau}$ for the Symmetron with non-minimal kinetic term and $\omega= 10^{10}$ in terms of dimensionless time $\tau$. The initial conditions are the same as that in Figure 2. } \label{fig4}
\end{figure}
Comparison of figures 3 and 4  indicates that in the Symmetron model, figure 3, variation of speed of the scalar field $\nu\equiv\frac{d\phi}{d\tau}$ is far greater in the time interval $0<\tau<0.002$ than in figure 4, where the Symmetron with non-minimal kinetic term is considered.

\section{Conclusions}
In this paper, we have proposed a Symmetron model with non-minimal kinetic term to increase the friction in the dynamical equation of motion of the scalar field. Since both the standard Symmetron and the proposed model behave in the same way in Newtonian and Post-Newtonian approximations, we conclude that local constraints remain unchanged and $m_{\phi}\geq H_{0}$.
By adding a non-minimal kinetic term to the action of the standard Symmetron we have shown that after symmetry breaking, positive acceleration will occur under specific conditions. In addition, unlike the Symmetron, no positive constant is needed in the potential $V(\phi)$. As a specific example, the evolution of the scalar field is numerically shown in figures 1 and 2 for the case of the standard Symmetron and  Symmetron with non-minimal kinetic term respectively. Comparison of the figures indicates that in spite of large $m_{\phi}$ in non-minimal kinetic model, the scalar field rolls down slowly due to more friction compared to the Symmetron model and as a result of the non-minimal kinetic term, the time evolution of the scalar field can be of the order of the Hubble time.

\end{document}